\documentclass[newstyle,twocolumn,proceedings]{rmaa}

\usepackage{rmaacite}

\newcommand{\msun}{\rm M_{\odot}}

\renewcommand{\P}[1]{%
\ifnum#1=1\hbox{OW~168--326E}\fi
\ifnum#1=2\hbox{OW~167--317}\fi
\ifnum#1=3\hbox{OW~163--317}\fi
\ifnum#1=5\hbox{OW~158--323}\fi
\ifnum#1=0\hbox{OW~171--334}\fi}

\title{The low-mass young stellar population in Orion OB1}
\author{C. Brice\~no,\altaffilmark{1}
  N. Calvet,\altaffilmark{2} A.K. Vivas,\altaffilmark{3} and L. Hartmann,\altaffilmark{2}}
\altaffiltext{1}{Centro de Investigaciones de Astronom{\'\i}a (CIDA), M\'erida, Venezuela.}
\altaffiltext{2}{Smithsonian Astrophysical Observatory (SAO), Cambridge, Massachusets, USA}
\altaffiltext{3}{Yale University, Astronomy Dept., New Haven, Connecticut, USA.}

\fulladdresses{
\item C. Brice\~no, CIDA, Apartado Postal
264, M\'erida 5101-A, Venezuela, (briceno@cida.ve).
\item N. Calvet, SAO, 160 Concord Av., Cambridge,
MA 02138, USA (ncalvet@cfa.harvard.edu).
\item A.K. Vivas, Yale University, Astronomy Department, P.O. Box 208101, New Haven, CT
06520-8101, USA (vivas@astro.yale.edu).
\item L. Hartmann, SAO, 160 Concord Av., Cambridge,
MA 02138, USA (lhartmann@cfa.harvard.edu).}

\shortauthor{BRICE\~NO ET AL.}
\shorttitle{The low-mass young stellar population in Orion OB1}

\keywords{Stars: formation, low-mass, pre-main sequence -- Surveys}

\abstract{%
We present recent results from our ongoing large scale variability
survey of the Orion OB1 Association.
In an area of $\sim 25$ square degrees we have unveiled new populations
of low-mass young stars over a range of environments, and
ages from 1-2 Myr in Ori OB 1b to roughly 10 Myr
in Ori OB 1a. There is a lack of strong H$\alpha$ emission and near-IR excesses 
in the young stars of Ori OB 1a, suggesting that accretion stops and
disks dissipate for most solar type stars in a few Myr,
probably caused by of the onset of planet formation.\\
The absence of dust and gas in Ori OB 1a is consistent with the 
picture of star formation as a rapid process, in which molecular
clouds dissipate in just a few Myr after the first stars are born.
}
\resumen{Versi\'on espa\~nol del ``abstract''}
\resumen{
Presentamos resultados recientes de nuestro sondeo de variabilidad
en la Asociaci\'on OB1 de Ori\'on.
En un \'area de $\sim 25$ grados cuadrados hemos develado nuevas poblaciones
de estrellas j\'ovenes de baja masa, en medios ambientes diversos y
abarcando edades desde 1-2 Ma\~nos en Ori OB 1b hasta aproximadamente
10 Ma\~nos en Ori OB 1a. La carencia de emisi\'on intensa en
H$\alpha$ y de excesos en el cercano IR en las estrellas j\'ovenes de Ori OB 1a,
sugiriendo que el acrecimiento de detiene y los discos se disipan en la 
mayor{\'\i}a de las estrellas tipo solar en cuesti\'on de unos Ma\~nos,
probablemente debido al inicio de la formaci\'on de planetas.\\
La ausencia de gas y polvo en Ori OB 1a es consistente con la idea 
que la formaci\'on estelar es un proceso r\'apido, en el que las nubes
moleculares se disipan en unos pocos Ma\~nos despu\'es de 
formadas las primeras estrellas.}

\listofauthors{C. Brice\~no, N.~Calvet, A.K.~Vivas, L.~Hartmann}
\indexauthor{Brice\~no, C.}
\indexauthor{Calvet, N.}
\indexauthor{Vivas, A.K.}
\indexauthor{Hartmann, L.}

\begin{document}

\maketitle

\section{Introduction}
\label{sec:intro}
During the last few years our understanding of the formation
of low-mass stars and planets has undergone major advances.
But still little is known about these processes in
in the vast areas spanned by nearby OB associations
like Orion OB1, were thousands of young, low-mass ($\la 1 \msun$)
stars are expected to exist but remain undetected.

To address this problem, we are carrying out 
a long-term optical variability survey spanning
$\sim 120 \> deg^2$ in the Orion OB1 Association ($d \sim 400$ pc),
to find, map, and study large numbers of widely-spread,
low mass ($\la 1 \msun$) stars with ages $\la$ 10 Myr.
The survey area (Figure 1) includes young regions of star
formation like the ONC ($\la 1$ Myr), the
Ori 1b sub-association (the belt region, $\sim 2$ Myr;
Warren \& Hesser [1977]; Brown et al. [1994]),
and older regions devoid of molecular gas like the Ori 1a 
sub-association ($\sim 11 $ Myr old).
\begin{figure}
  \hskip1.0cm\includegraphics[width=5.9cm]{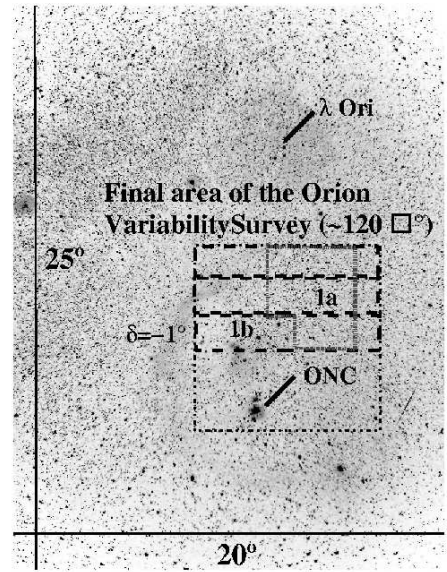}
  \vskip -0.3cm
  \figcaption{Image of Orion showing the total survey area of $120\> deg^2$
    (large dashed-lined box). The initial strip at $\delta=-1^o$, 
    passing over the three Orion belt stars (Ori 1b) and part of Ori 1a
    is indicated.
    Two addditional strips covering the northern part of our survey
    have also been completed. The Orion Nebula Cluster and the
    bubble around the star $\lambda$ Orionis are clearly seen.}
  \label{fig-1}
\end{figure}

\bigskip
\section{The Variability Survey: \\ initial results}
\label{sec:surv}
The multiband (BVRIH$\alpha$), multi-epoch, survey
is being carried out using an 8k x 8k
pixel CCD Mosaic Camera (Baltay et al. 2002)
installed on the 1.0/1.5m
Schmidt telescope at The National Astronomical 
Observatory of Venezuela ($8^\circ 47'$ N, 3610 m elevation).
The camera is optimized for drift-scanning,
generating a continuous $2.3^\circ$ wide strip 
of the sky at a rate of $34.5 \> deg^2/hr/filter$,
down to $V_{lim}= 19.7$ ($S/N=10$).  
We identify variable stars using a $\chi^2$ test at a 99.99\%
confidence level.  

Among the bright objects ($V\la 16$) 
in a strip centered at $\delta=-1^o$ (Figure 1),
we selected candidate variable stars located
above the zero age main sequence (ZAMS) in a $V$ vs. $V-I$ diagram.
Followup spectroscopy was obtained 
using the FAST spectrograph on the 1.5m telescope of the Smithsonian 
Astrophysical Observatory at Mt. Hopkins, Arizona, with a spectral 
resolution of 6.5{\AA } over the range 4000 - 7000\AA. 
Of 350 candidates, 180 were confirmed as low mass pre-main sequence 
stars (T Tauri stars - TTS), based on the presence of
emission lines such as H$\alpha$ and the absorption line Li I 6707{\AA }
(an indicator of youth in late type stars [Brice\~no et al. 1997, 1999]).
The newly identified TTS have spectral types K3 - M2 
($M_* \sim 0.9-0.6\>M_{\odot}$).

Figure 2 shows color-magnitude diagrams for stars in Ori 1a and 1b. 
The data for each star are median
values determined from the multiple observations of each object.
It is apparent that stars in 1a are older than stars in 1b.
Stars in 1b seem to fall between the isochrones
corresponding to 1 - 3 Myr, while stars
in 1a fall between 3 and 30 Myr.

Using the near-IR $JHK$ data for
the new TTS, obtained from the second release
of the 2 Micron All Sky Survey (2MASS), we find that
all stars in Ori 1a are within the region expected
for purely stellar emission;
in contrast, many stars in 1b have large
$H-K$ colors indicative of excess emission from a hot inner dusty disk,
and exhibit the strong H$\alpha$ emission (W[H$\alpha] \ga 10\AA$)
associated with accretion from a circumstellar disk onto the central star.
The lack of H$\alpha$ emission in Ori 1a
stars indicates that protoplanetary
disk accretion stops for almost all solar-type stars; the absence of
near-IR excesses shows that significant inner disk dissipation ocurrs
in a few Myr, possibly caused by coagulation of the dust particles
into larger bodies like planetesimals/planets.

Our results have important implications
for the  star forming history of the region.
We see a well-defined older association, Ori 1a,
resembling a fossil version
of the younger Ori 1b located next to it,
a plausible scenario for triggered or sequential star formation.
Also, the absence of molecular gas in Ori 1a
supports suggestions that large molecular cloud complexes
can form stars and disperse in only a few Myr
(Ballesteros et al. 1999).
\begin{figure}
  \includegraphics[width=\columnwidth]{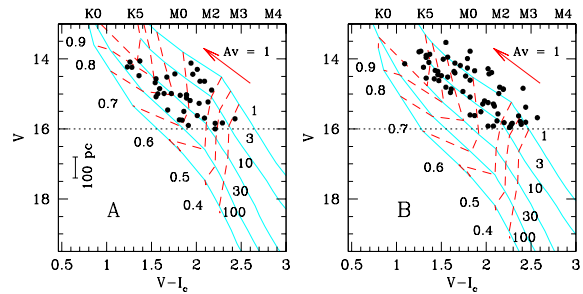}
  \vskip -0.7cm
    \caption{ $V$ vs. $(V-I_c)$ diagram for the brighter new TTS in
      Ori 1a (A; d=330 pc) and Ori 1b (B; d=460 pc). Isochrones (solid lines)
      for ages 1 to 100 Myr and evolutionary tracks (dashed lines) for
      masses 0.4 to 0.9 $\msun$ are indicated (Baraffe et al. 1998). 
      The shifts due to 1 magnitude of de-reddening (arrow)
      and to a distance change of 100 pc (left vertical bar)
      are also indicated. The dotted lines show the V=16 limit
      of FAST spectroscopy.}
  \label{fig-2}
\end{figure}

\acknowledgments

We acknowledge support from National Science Foundation (NSF) grant AST-9987367.
Research reported herein based on observations made with
the 1m Schmidt Telescope at Observatorio Astron\'omico
Nacional de Llano del Hato, M\'erida, Venezuela, 
operated by CIDA and funded by the Ministerio de Ciencia y
Tecnolog{\'\i}a and the Fondo Nacional de Ciencia y Tecnolog{\'\i}a
of Venezuela.
This publication makes use of data products from the Two Micron All Sky
Survey, a joint project of the University of Massachusetts and the
Infrared Processing and Analysis Center/California Institute of Technology,
funded by NASA and the NSF.

\end{document}